\documentclass[graphicx]{article}

\usepackage{epsfig}

\def\be{\begin{eqnarray}}
\def\ee{\end{eqnarray}}
\def\nn{\nonumber}

\def\({\left(}
\def\){\right)}
\def\labels#1{\label{#1}}
\def\eq#1{(\ref{#1})}
\def\G{{\cal G}}

\def\fig#1{Fig.~\ref{fig:#1}}
\def\tab#1{Table~\ref{table:#1}}
\def\bk#1{\langle#1\rangle}

\begin{document}

\title{Invariant and Group Theoretical Integrations over the $U(n)$ Group}
\author{S.~Aubert and C.\,S.~Lam \\
Department of Physics, McGill University \\
3600 University St., Montreal, QC, Canada H3A 2T8 \\
Emails: samuel.aubert@elf.mcgill.ca, lam@physics.mcgill.ca}
\date{}
\maketitle

\begin{abstract}
In a previous article, an `invariant method' to calculate monomial
integrals over the $U(n)$ group was introduced. In this paper, we
study the more traditional group-theoretical method, and compare its
strengths and weaknesses with those of the invariant method. As a
result, we are able to introduce a `hybrid method' which combines the
respective strengths of the other two methods. There are many examples
in the paper illustrating how each of these methods works.
\end{abstract}

\section{Introduction}

This article deals with the calculation of integrals of the form
\begin{equation}
\int (dU)\, U_{i_1 j_1}^* \cdots U_{i_p j_p}^* U_{k_1 l_1}
\cdots U_{k_q l_q} \labels{intGeneral}  
\end{equation}
over the $U(n)$ group, where $(dU)$ is the invariant Haar measure
normalized to $\int (dU) = 1$, and $U_{ij}$ is a $U(n)$ matrix
element, with $U^*_{ij}$ being its complex conjugate. 

These integrals and  their generating functions are useful in many
areas of physics, including two-dimensional
quantum gravity \cite{DGZ-J}, QCD, matrix models, and
statistical and condensed-matter problems of various
sorts \cite{GM-GW}. They are also needed in
the parton saturation problem at small Feynman-$x$ \cite{LMZ}.

The integral \eq{intGeneral} depends on the indices $I=\{i_1\cdots i_p\},
J=\{j_1\cdots j_p\},K=\{k_1\cdots k_q\}$, and $L=\{l_1\cdots l_q\}$,
so it will be denoted as $\bk{IJ|KL}$:
\be
\bk{IJ|KL}=\int (dU)\, U^*_{IJ}U_{KL},\labels{intGeneral2}
\ee
where $U^*_{IJ}=\prod_{a=1}^pU^*_{i_aj_a}$, and similarly for $U_{KL}$.
Since the matrix elements commute,
$U^*_{IJ}=U^*_{I_PJ_P}$
where $I_P=\{i_{P(1)}\cdots i_{P(p)}\}$ is obtained from $I$
by a permutation $P\in S_p$ of its $p$ indices. Hence
\be
\bk{IJ|KL}=\bk{I_PJ_P|K_TL_T}\labels{perm}\ee
for any $P\in S_p$ and $T\in S_q$.

The integral is nonzero only when $p=q$, a number which will be
referred to as the {\it degree} of the integral. Without loss of
generality, it turns out that we may assume $K=I$ and $L$ to be a
permutation of $J$, namely, $L=J_Q$ for some $Q\in S_p$. The value of the integral depends on what the index sets
$I,J$, and what the element $Q$ are, so even for a given $p$, 
there are many distinct
cases. The best way to distinguish them is to represent each integral
by a diagram in a way to be explained in the next section. 

Integral \eq{intGeneral} has been computed using a graphical
technique \cite{Cr}. It can also be obtained using
the Itzykson-Zuber formula \cite{IZ} as a generating function,
or directly from group theory \cite{GT}
using the Frobenius formula \cite{Weyl}.
We shall refer to this last method as the {\it group-theoretical method},
or GTM for short. In the GTM, a general
formula is available to compute \eq{intGeneral}. It involves a triple sum over an
expression containing characters of the symmetric group $S_p$, as well
as the dimensions of irreducible representations of $S_p$ and of the
unitary group $U(n)$. One of the sums is taken over all the relevant
irreducible representations, and the others are taken over the symmetry
groups of the index sets $I$ and $J$.
These sums could be long and tedious for a large $p$, and for 
most symmetry groups.

A different way to calculate \eq{intGeneral} was introduced in a recent
paper \cite{AL}. This method relies only on the unitary nature of the
matrix elements, and the invariance of the Haar measure. In particular,
no knowledge of group theory is necessary. The invariance of the Haar
measure, as well as the off-diagonal unitarity relation, are
used to derive relations between integrals of the same degree.
The diagonal unitarity relation connects integrals of degree $p$ with
ones of degree $p-1$. Through a chain of these relations, the desired
integral is finally related to the basic integral of degree 0, which is
$\int dU=1$. The desired integral is then  solved from this chain
of relations. We have called this method the {\it invariant method},
or IM for short. 

The purpose of this paper is to compare the pros and cons of the
GTM with the IM. In order to do so we must
first study and understand better the nature
of the GTM. Armed with this comparison, we will be able to design
a new {\it hybrid method} which combines the strengths of these
two other methods.

The IM is reviewed in Sec.~2. It is used to derive a new 
`double-fan' relation needed in the example for the hybrid method.

The GTM is reviewed and studied in Sec.~3. It can be used to derive
simple relations, but the relations derived in this way are nowhere as
powerful as those derived with the IM. The group-theoretical formula
can be used to calculate any integral, but generally that is tedious
and has to be done integral by integral. However, for integrals whose
two symmetry groups are disjoint, or one is contained in the other,
systematics emerge to make the calculation simpler.
Several of these `orderly' integrals are studied in Sec.~3. 

A comparison of the strengths and weaknesses of the two methods
is to be found in Sec.~4. Armed with an understanding of their
relative merits, we design a `hybrid method' in Sec.~5 to take
advantage of their respective strengths. This method is 
illustrated by the calculation of a class of `double-fan' integrals.
There are also four Appendices showing details of various calculations.

\section{The Invariant Method (IM)}
\subsection{A Brief Review} \label{sec:IM}

The invariant method presented in a previous
paper~\cite{AL} can be used to calculate the integral
\eq{intGeneral}. The method exploits the unitarity of the $U(n)$
group elements,   
\begin{equation} 
\sum_{j=1}^n U_{ij}^*U_{lj} = \sum_{j=1}^n U_{ji}^*U_{jl} =
\delta_{il}, \labels{unitarity} 
\end{equation} 
and the invariance of the Haar measure, in the form 
\begin{eqnarray}
\int (dU)\, f(U,U^*) 
&\!=\!& \int (dU)\, f(U^*,U) = \int (dU)\, f(U^T,U^{*\,T}) \nn \\  
&\!=\!& \int (dU)\, f(VU,V^*U^*) = \int (dU)\, f(UV,U^*V^*), 
\label{inv} \end{eqnarray}
for any function $f$ and any $V \in U(n)$.  

The unitarity relation (\ref{unitarity}) relates integrals of the same
degree if $i \not= l$, and it relates integrals of degree $p$ to 
integrals of degree $p-1$ if $i=l$. Other relations between integrals
of the same degree can be obtained from (\ref{inv}), by suitable choices
of $V$. Here are some of them discussed in the previous paper
\cite{AL}.     
  
\begin{enumerate} 
\item Using $V_{ij} = e^{i \phi}\delta_{ij}$, it follows that $p$ must
be equal to $q$ in order to avoid the vanishing of integral
(\ref{intGeneral2}). The number $p$ will be called the {\it degree} of
the integral. 

\item Using $V_{ij} = e^{i \phi_i} \delta_{ij}$, it follows that
(\ref{intGeneral2}) is nonzero only when $K=I_M$ and $L=J_R$ for some
$M,R \in S_p$. Using \eq{perm}, and denoting $RM^{-1}$ by $Q$, only
integrals of the type 
\begin{equation}
\langle IJ|IJ_Q \rangle = \int (dU)\, U_{IJ}^*\, U_{IJ_Q} 
\labels{intSimple}
\end{equation}
are nonzero, so from now on we need  to consider only integrals
of this type. 

An integral with $J_Q=J$ will be called a {\it direct integral}. For
such integrals we may always choose $Q=e$, the identity permutation.
Otherwise, the integral is an {\it exchange integral}.

Integrals are represented diagrammatically as follows. Each
distinct value in the index set $I$ is represented by a dot on the
left (L-dot), and each distinct value of the index set $J$ or $J_Q$
is represented by a dot on the right (R-dot). The factor $U^*_{ij}$ is
shown as a thin solid line between the L-dot $i$ and the R-dot $j$,
and the factor $U_{ij}$ is shown as a dotted line between these two
dots. The factor $U^*_{ij}U_{ij}$ is represented by a thick line, or
more generally, the factor ${U^*}_{ij}^mU_{ij}^n$ is represented
by a thick line with a pair of numbers $(m,n)$ written beside it. If
$m=n$, then only a single number $m$ is written. The numbers $(m,n)$
or $m$ will be known as the {\it multiplicities} of the line. See
Fig.~\ref{fig:excZ} for an illustration.  

\begin{figure} \begin{center}
\includegraphics{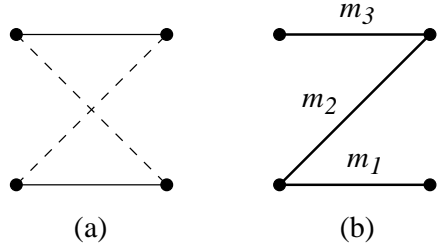}  
\caption{Examples of $U(n)$ integral diagrams. (a)~The unique
exchange integral for $p = 2$; (b)~a $Z$-integral with arbitrary
multiplicities $m_1$, $m_2$, and $m_3$.} 
\labels{fig:excZ} 
\end{center} \end{figure}

\item With $V$ chosen to be a permutation matrix of {\it n} objects,
it follows that  
\[
\langle IJ| KL \rangle = \langle I'J| K'L \rangle = 
\langle IJ'| KL' \rangle, 
\] 
where $I'$ is obtained from $I$ by a reassignment of the values of its
indices, e.g.\ if $I=(334)$, then $I'$ may be $(558)$. $K'$ is
obtained from $K$ by the {\it same} reassignment, and similarly for
$J'$ and $L'$. As a result, there is no need to know the values of
the indices of the L-dots, nor the R-dots.  This is why the dots
in \fig{excZ} are not labelled.

\item As a consequence of the first two equalities in (\ref{inv}), an
integral remains the same under the interchange of the solid lines
with the dotted lines, or the L-dots with the R-dots. 

\item Using $V = R(ab)$, the rotation matrix in the $(a,b)$
plane, a `spin-off relation' is obtained. Consider a R-dot imbedded in
an arbitrary integral $M_0 = \langle IJ|IJ_Q \rangle$, with $d$ pairs of
solid-dotted lines attached to the dot. Now spin off $e$ pairs of
these lines to create a new R-dot and a new integral. There are many
ways to choose the $e$ pair of lines, each possibly corresponds to a
different integral. Let $M_e$ be the sum of all these integrals. Then
the quantities $M_0$ and $M_e$ are related by the spin-off relation 
\begin{equation} 
M_e = M_0\, \left( \!\! \begin{array}{c} d \\ e \end{array} \!\!
\right), \labels{rotation} 
\end{equation} 
where $\left( \!\! \begin{array}{c} d \\ e \end{array}\!\! \right)$ is
the binomial coefficient. 

The relation is local in that it is independent of the the structure of the
rest of the diagram.  
The same relation can also be used to spin off a L-dot.
\label{item:rotation}   
\end{enumerate}

In what follows, we summarize two general results obtained in the
previous paper \cite{AL} using the IM.

\subsubsection{The fan relation}

\begin{figure} \begin{center}
\epsfig{file = 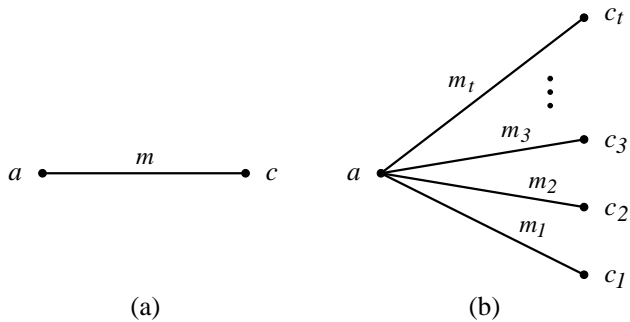, scale=0.9}  
\caption{The fan diagrams shown here can be a part of a larger
diagram. In that case, there may be many more dots and many more lines
in the complete diagram, provided none of the additional lines land on
the R-dots shown. (a) A closed fan; (b) a partially opened fan. If all
$m_i=1$, then it is said to be a fully opened fan, or just an opened
fan.} \labels{fig:originalFan} 
\end{center} \end{figure}

The {\it fan relation} 
\begin{equation}
\int (dU)\, A\, |U_{ac_1}|^{2m_1} \cdots |U_{ac_t}|^{2m_t}\,
= \frac{(\prod_{j=1}^t m_j!)}{(\sum_{j=1}^t m_j)!} 
    \int (dU)\, A\, |U_{ac}|^{2m}
\labels{fan}   
\end{equation}
relates integrals of the same $m=\sum_{j=1}^t m_j$, where $A$ is an
arbitrary product of matrix elements of $U$ and $U^*$ whose column
indices are different from $c_1,c_2, \ldots ,c_t$. The column index
$c$ on the right could be taken to be one of the $c_i$'s.

Diagrammatically, the integral on the right of \eq{fan} is shown in
Fig.~\ref{fig:originalFan}(a), and the integral on the left is shown
in Fig.~\ref{fig:originalFan}(b). The additional lines and dots
corresponding to the factor $A$ are not shown, because they do not
affect the spin-off relation. We shall refer to
Fig.~\ref{fig:originalFan}(a) as a {\it closed fan}, and
Fig.~\ref{fig:originalFan}(b) as a {\it partially opened fan}. If
every $m_i=1$, then it will be said to be a {\it fully opened fan}, or
simply an {\it opened fan}.

In particular, a closed fan integral is $m!$ times an opened fan
integral. In fact, this relation between the two types of fans
immediately gives rise to the relation \eq{fan} between a closed fan
and a partially opened fan. To see it, note that each branch of a
partially opened fan is itself a closed fan. By opening up all of
them, we get the fully opened fan integral, multiplied by a
multiplicity factor $\prod_j m_j!$ from all the branches. Thus a
closed fan is $m!/\prod_j m_j!$ times a partially opened fan, as given
by \eq{fan}. 

\subsubsection{The $Z$-integral and the fan integral}

In \cite{AL}, we also obtained a general formula for integrals of the
type shown in Fig.~\ref{fig:excZ}(b), for arbitrary nonnegative
integers $m_1,m_2$, and $m_3$. We call that the `$Z$-formula' because of
the shape of the graph. It is 
\begin{eqnarray} 
Z(m_1,m_2,m_3)\hspace{-3mm}&\equiv&\hspace{-3mm}\int (dU)\,|U_{ij}|^{2 m_1}
 |U_{il}|^{2m_2} |U_{kl}|^{2m_3} \nonumber \\ 
 &=&\hspace{-3mm}\frac{m_1!\, m_2!\, m_3!\, (n-2)!(n-1)!\, (n+m_1+m_3-2)!}{(n+m_1-2)!\,
(n+m_3-2)!\, (n+m_1+m_2+m_3-1)!}. \labels{Z}  
\end{eqnarray}

In the special case where $m_2=m_3=0$, this becomes
\be
F(m) \equiv Z(m,0,0) = {m!(n-1)!\over(n+m-1)!}, \labels{zf}
\ee
which is the integral for Fig.~\ref{fig:originalFan}(a) when there are
no additional dots or lines around. 

\subsection{Double-Fan Relation}

\begin{figure} \begin{center}
\includegraphics{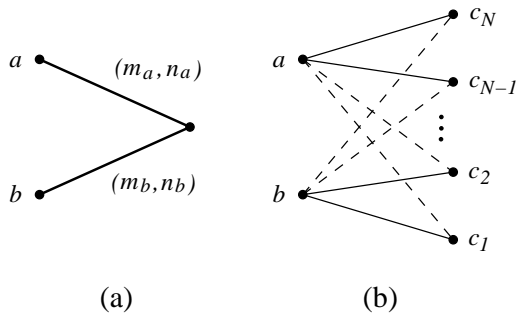}  
\caption{Double-fan diagrams. (a) A closed diagram; (b) a fully opened
diagram.}   
\labels{fig:compound2} 
\end{center} \end{figure}

The fan relation can be generalized to a double-fan relation, connecting
 the closed `double-fan' diagram of Fig.~\ref{fig:compound2}(a)
with a (fully) opened double-fan diagram such as
Fig.~\ref{fig:compound2}(b). As in Fig.~\ref{fig:originalFan}, there
may be additional dots and lines in the integral, but none of
them may end up on the R-dots shown. From that relation, we
can also deduce relations between a closed
double-fan and a partially opened double-fan, as done in the single-fan
case.  

The double-fan relation is considerably more complicated than the 
single-fan relation \eq{fan}, because there are many more double-fan
graphs. Each R-dot of a (fully) opened (double-fan) graph such as
Fig.~\ref{fig:compound2}(b) falls into one of four {\it basic patterns}:
$[A_a],[A_b], [B_a]$, and $[B_b]$, shown in
Fig.~\ref{fig:basic}. If the solid and dotted lines end up on the same
L-dot, the pattern is a $[B]$; otherwise it is an $[A]$. The
subscripts $a$ and $b$ tell us which L-dot the {\it solid} line
emerges from. 

Suppose there are $\alpha_i$ number of $[A_i]$ and $\beta_i$ number of
$[B_i]$ patterns in a (fully) opened (double-fan) graph. Then there
are $m_a$ solid and $n_a$ dotted lines emerging from the L-dot $a$,
and  $m_b$ solid and $n_b$ dotted lines emerging from the L-dot $b$,
where 
\be
m_a &\!\!=\!\!& \alpha_a + \beta_a, \qquad
n_a \;=\; \alpha_b + \beta_a, \nn \\
m_b &\!\!=\!\!& \alpha_b + \beta_b,\qquad\,
n_b \;=\; \alpha_a + \beta_b. \labels{abmn}
\ee 
The total number of R-dots in the opened graph is $N=m_a+m_b=n_a+n_b$.   

\begin{figure} \begin{center}
\includegraphics{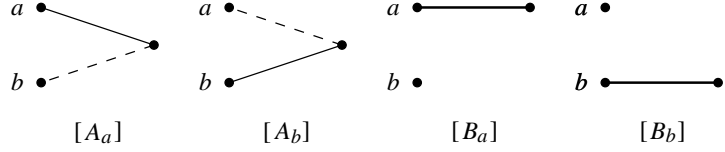}  
\caption{The four basic patterns for the R-dots of a fully opened
double-fan graph.} \labels{fig:basic}  
\end{center} \end{figure}

When these $N$ R-dots are merged together, we get the closed
(double-fan) graph depicted in Fig.~\ref{fig:compound2}(a), which will be
denoted by $[(m_an_a)(m_bn_b)]$. If $N=1$, this is just one of the
four basic patterns discussed before. If $N>1$, we will call it a {\it
compound pattern}. 

From \eq{abmn}, we see that if we replace $\alpha_i$ and $\beta_i$ by
$\alpha_i'=\alpha_i+\xi$ and $\beta_i'=\beta_i-\xi$, with any
integral $\xi$ which keeps $\alpha_i'$ and $\beta_i'$ nonnegative,
then we get the same closed graph by collapsing this new opened graph.
Conversely, it will be shown in Appendix~\ref{sec:eqns_VV_v} that the
closed graph $[(m_an_a)(m_bn_b)]$ can be spinned off into a sum of
several opened graphs, one for each $(\alpha_i',\beta_i')$. The
double-fan relation expressing that quantitatively is 
\be
[(m_an_a)(m_bn_b)]&=& \sum v(\alpha_a' \alpha_b' \beta_a'
\beta_b')\, [A_a]^{\alpha_a'} [A_b]^{\alpha_b'}
[B_a]^{\beta_a'} [B_b]^{\beta_b'}\nn\\ 
&\equiv&\sum[\alpha_a'A_a+\alpha_b'A_b+\beta_a'B_a+\beta_b'
B_b], \labels{VV} \ee where 
\be
v(\alpha_a'\alpha_b' \beta_a'\beta_b') 
&=& {m_a!\,n_a!\,m_b!\,n_b! \over
\alpha_a'!\,\alpha_b'!\,\beta_a'!\,\beta_b'!},  
\labels{v} \ee
and the sum is over all solutions
$(\alpha_a',\alpha_b',\beta_a',\beta_b')$ 
of \eq{abmn}. 

The double fan becomes a single fan if the L-dot $b$ is not connected,
namely, if $m_b=n_b=0$ and $m_a=n_a \equiv m$. In that case \eq{VV}
becomes  
\[
[m\, B_a] = m!\ [B_a]^m,
\labels{sfan} \]
which is just \eq{fan} (when all $m_i=1$) in another notation.

\section{Group-Theoretical Method (GTM)} \label{sec:group}
\subsection{A Brief Review}

Using group theory, the integral (\ref{intSimple}) can be
turned into a multiple sum \cite{IZ,GT}. In the notation used
in Appendix A of \cite{AL}, the formula is  
\begin{equation}
\langle IJ|IJ_Q \rangle = \sum_{R \in {\cal G}_I}\, \sum_{S \in
{\cal G}_J}\, \sum_f \frac{d_f^2}{(p!)^2\, \tilde{d}_f}\, \chi_f(SQR),
\labels{intGT} 
\end{equation}
where $p$ is the degree of the integral. The symbols ${\cal G}_I$ and
${\cal G}_J$ represent the symmetry groups of the row and column index
sets. More precisely, ${\cal G}_X = \{ P \in S_p | P(X) = X \}$, with
$X$ being either $I$ or $J$. The irreducible representations of the
symmetric and unitary groups are both labelled by a sequence $f =
(f_1,f_2, \ldots ,f_p)$, with $f_1 \geq f_2 \geq \cdots \geq f_p \geq
0$. $\chi_f(P)$ is the character of $P \in S_p$ in the irreducible
representation with signature~$f$. The dimension of the irreducible
representation $f$ is given by $d_f = \chi_f(e)$ for $S_p$, and by
$\tilde{d}_f$ for $U(n)$. A formula for $\tilde{d}_f$ is: 
\begin{equation} \tilde{d}(f_1, \ldots, f_n) = \frac{D(f_1 +
(n-1), f_2 + (n-2), \ldots, f_n)}{D(n-1,n-2, \ldots, 0)},
\labels{dUn} \end{equation} 
where $D(x_1, \ldots, x_n)$ is the Vandermonde determinant given by
$ \prod_{i<k} (x_i - x_k).$ 

Since $\chi_f(g)$ depends only on the class $c_g$ that $g$ belongs to,
we may write it as $\chi_f(c_g)$. With this notation,
eq.~(\ref{intGT}) can be re-written  as
\begin{equation}
\langle{IJ|IJ_Q}\rangle = \sum_c N[c]\, \xi[c], \labels{intPrim}
\end{equation} 
where the sum is taken over all classes $c$ of $S_p$,
\begin{equation}
N[c] = \sum_{R \in {\cal G}_I} \sum_{S \in {\cal G}_J}
\delta(SQR \in c) \labels{N} 
\end{equation}
is the number of elements of the type $SQR$ in the class $c$, and
\begin{equation}
\xi[c]= \sum_f \frac{d_f^2}{(p!)^2\, \tilde{d}_f} \chi_f(c).
\labels{xi} 
\end{equation}

It is not difficult to see that $Q$ is not unique, because $Q'=SQ=QT$ 
for any $S\in{\cal G}_J$ and any $T\in{\cal G}_{J_Q}$
is another possible $Q$. It does not matter which $Q$ we pick in
\eq{N}. That equation can also be written as
\be
N[c] = \sum_{R \in {\cal G}_I} \sum_{T \in {\cal G}_{J_Q}}
\delta(QTR \in c). \labels{Nq}\ee

It is straightforward but
generally very tedious to compute $N[c]$, because we
need to calculate the product 
$QTR$ for every $T\in{\cal G}_{J_Q}$,
every $R\in{\cal G}_I$, and determine 
what class $c$ the product belongs to. Then we have to count
up all the products that are in a given class $c$ to get $N[c]$. 
However, the task becomes considerably more manageable 
 if either ${\cal G}_I$ and
${\cal G}_{J_Q}$ are disjoint, or if one is contained in the other.
We shall refer to integrals with those properties as {\it orderly}.
Further simplification occurs for 
direct integrals, because in that case  $Q$ can always
be chosen to be the identity $e$, so the triple product
is reduced to a double product $TR$. 

The calculation of $\xi[c]$ in \eq{xi} is 
simpler than the calculation of $N[c]$, but still we know of  no
closed form of it
 valid for every class $c$ and every symmetric group
$S_p$. The best we can do is to compute them case by case.
Results are given in Sec.~3.3.
Each $\xi[c]$ is actually an orderly integral with $\G_I=\G_J=e$,
to be referred to as a {\it primitive integral}.

Other integrals
can be computed in terms of the primitive integrals, if $N[c]$
is known. We shall discuss two orderly integrals for which
$N[c]$ can easily be obtained. In Sec.~3.4, we discuss the 
{\it stack integrals}, which are
direct integrals with $\G_I=\G_J$. In Sec.~3.5, we discuss
the fully opened double-fan integrals of the type $[A_a]^\alpha
[A_b]^\alpha$.

Relations between orderly integrals
may be obtained without knowing the explicit values of $\xi[c]$, 
if their $N[c]$'s are related in a simple way. This is the case
for the single-fan relation, and the double-fan relation 
with $n_a=m_b=0$ and $m_a=n_b=m$. They will be discussed in
Sec.~3.2. However, general double-fan integrals are not orderly,
so we cannot obtain the general double-fan relation by the
group-theoretical method, at least not in the present way.
We will also show that the closed (single-)fan
{\it integral} can also be computed without explicitly knowing what 
$\xi[c]$ are. This is one of the very few cases where
integrals can be obtained group-theoretically without 
explicitly knowing $\xi[c]$.

That leaves the non-orderly integrals, for which each term of
the summand in \eq{Nq} has to be calculated separately to get
$N[c]$. The first non-orderly integral occurs in degree $p=3$.
In Sec.~3.6, we shall show how to calculate some of the 
$p=3$ and $p=4$ non-orderly integrals.

\subsection{Single-Fan and Simple Double-Fan Relations}
The single-fan integrals are orderly. The index sets
for the closed fan \fig{originalFan}(a) are
\be
\pmatrix{label\cr I\cr J\cr J_Q\cr}&=&
\pmatrix{1&\cdots&m&\cdots\cr
a&\cdots& a&\cdots\cr
c&\cdots&c&\cr
c&\cdots&c&\cr
}.\labels{tab1}\ee
The first row gives the index labels, and the next three rows give
the values of the indices in the sets $I,J,$ and $J_Q$ respectively.
Different letters are understood to correspond to different values. 
Additional dots and lines may be
present in the graph, as long as none of the lines end up in the
R-dots shown. These additional lines and dots are not drawn because they do not
affect the fan relation in any way. Similarly, they are not shown
in the index sets in  \eq{tab1} other than the ellipses in the first two rows,
which remind us that there may be more lines connected to the L-dots.
 Such ellipses are
absent in the last two rows because no additional lines are allowed to 
be connected to the R-dots shown. 

Similarly, the index sets for the fully opened fan, \fig{originalFan}(b)
with all $m_i=1$, are 
\be
\pmatrix{label\cr I\cr J\cr J_Q\cr}&=&
\pmatrix{1&\cdots&m&\cdots\cr
a&\cdots& a&\cdots\cr
c_1&\cdots&c_m&\cr
c_1&\cdots&c_m&\cr
}.\labels{tab2}\ee

Using $S_m$ to denote the symmetric group for the permutation
of the first $m$ labels, the symmetric groups for \fig{originalFan}(a)
can be read off
from \eq{tab1} to be $\G_I\supset S_m$ and $\G_J=\G_{J_Q}=S_m$.
Similarly, the symmetric groups for \fig{originalFan}(b)
can be read off
from \eq{tab2} to be $\G_I\supset S_m$ and $\G_J=\G_{J_Q}=e$.
We may choose $Q=e$ in both cases. Then 
for \fig{originalFan}(a), $T{\cal G}_I={\cal G}_I$
for every $T\in{\cal G}_{J_Q}$, hence $N[c]=m!\sum_{R\in\G_I}
\delta(R\in c)$. But the last sum is simply the $N[c]$
for \fig{originalFan}(b) and \eq{tab2}. Hence 
it follows from
\eq{intPrim} that the fan relation (with all $m_i=1$) is true.

Next, let us derive the double-fan relation \eq{VV} and \eq{v}
for the case $n_a=m_b=0$ and $m_a=n_b$. The solution of
\eq{abmn} is now unique. It gives $\alpha_a=m_a=n_b\equiv m$, and
$\alpha_b=\beta_a=\beta_b=0$. The double-fan
relation \eq{VV} then becomes
\be
[m\, A_a]=m!\,[A_a]^m.\labels{fansd}\ee

The closed double-fan is shown
in \fig{fan}(a). Its index sets are
\be
\pmatrix{label\cr I\cr J\cr J_Q\cr}&=&
\pmatrix{1&\cdots&m&\cdots&n&\dots&n+m&\cdots\cr
a&\cdots&a&\cdots&b&\cdots&b&\cdots\cr
c&\cdots&c&&&&&\cr
&&&&c&\cdots&c&\cr}.\labels{tab3}\ee
The opened double-fan is shown
in \fig{fan}(b). Its index sets are
\be
\pmatrix{label\cr I\cr J\cr J_Q\cr}&=&
\pmatrix{1&\cdots&m&\cdots&n&\dots&n+m&\cdots\cr
a&\cdots&a&\cdots&b&\cdots&b&\cdots\cr
c_1&\cdots&c_m&&&&&\cr
&&&&c_1&\cdots&c_m&\cr}.\labels{tab4}\ee
We may choose $Q=(1,n)(2,n+1)\cdots(m,n+m)$
in both cases. For \eq{tab3},
$\G_I\supset S_m\otimes S'_m,\G_J=S_m$,
and $\G_{J_Q}=S'_m$, where $S_m$ is the
permutation group of the first $m$ labels,
and $S_m'$ is the permutation group for
the labels $(n,n+1,\cdots,n+m)$. 
For \eq{tab4},
$\G_I\supset S_m\otimes S'_m$, but $\G_J
=\G_{J_Q}=e$.

For \fig{fan}(a), $T\G_I=\G_I$
for every $T\in\G_{J_Q}$. Hence
$N[c]=m!\sum_{T\in\G_I}\delta(QR\in c)$.
But the last sum is simply the $N[c]$
of \fig{fan}(b). In this way \eq{fansd}
is proven by the GTM. 

The fan {\it integral} \eq{zf} can also be obtained
from the GTM. 
It is given by \fig{originalFan}(a) without extra dots and lines,
or \eq{tab1} without the ellipses at the end.
Then $\G_I=S_m$, and $N[c]=m!\sum_R\delta(R\in c)$.
From \eq{intPrim} and \eq{xi}, we get
\be
\bk{IJ|IJ_Q}&\equiv&F(m)=
m!\,
\sum_{R\, \in\, {\cal G}_I} \sum_f \frac{d_f^2}{(m!)^2\,
\tilde{d}_f}\, \chi_f(R) \labels{first} \\ 
& = & \frac{1}{m!}\, \sum_f
\frac{d_f^2}{\tilde{d}_f} \sum_{R\, \in\, S_m} \chi_f(R)\:
\chi_{(m)}^*(R)\, =\,
 \frac{d_{(m)}^2}{\tilde{d}_{(m)}} \labels{third} \\
& = & \frac{m!(n-1)!}{(m+n-1)!}. \labels{fourth}
\end{eqnarray}
In getting from \eq{first} to \eq{third}, the character
$\chi_{(m)}^\ast(R)=1$ of the totally symmetric representation
$(m)$ of the permutation group has been inserted, and the
orthogonality relation of the characters has been used.
To get to \eq{fourth}, $d_{(m)}=\chi_{(m)}(e)=1$ as well as
$\tilde d_{(m)}=(m+n-1)!/(n-1)!m!$ (see \eq{dUn}) have \nobreak been used.

The result in \eq{fourth} agrees with the result \eq{zf}. It is one of the
very few cases where the value of the integrals can be obtained group-theoretically
without knowing the values of the individual $\xi[c]$'s.

\begin{figure} \begin{center}
\includegraphics[scale=0.9]{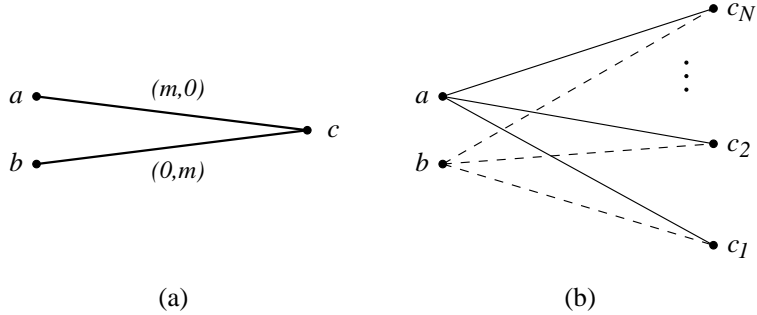}  
\caption{Double-fan diagrams with $n_a=m_b=0$, and
$m_a=n_b=m$. (a) A closed fan; (b) a fully opened
fan.} \labels{fig:fan}  
\end{center} \end{figure}

\subsection{Primitive Integrals} \labels{sec:primitive}
Integrals in which both symmetry groups ${\cal G}_I$ and ${\cal G}_J$
consist only of the identity $e$ will be called {\it primitive}. This
happens when all the indices $i_a$ in the set $I$ assume distinct
values, and all the indices $j_b$ in the set $J$ are also
different. The corresponding diagrams have $p$ dots each on both
columns, and precisely one solid and one dotted lines connecting to
each of the dots. The primitive diagrams for $p \le 3$ are shown in
Fig.~\ref{fig:prim}, and the ones for $p = 4,5$ are contained in
Appendix~\ref{sec:prim45}.   

\begin{figure} \begin{center}
\vspace{1cm}
\includegraphics[scale=0.9]{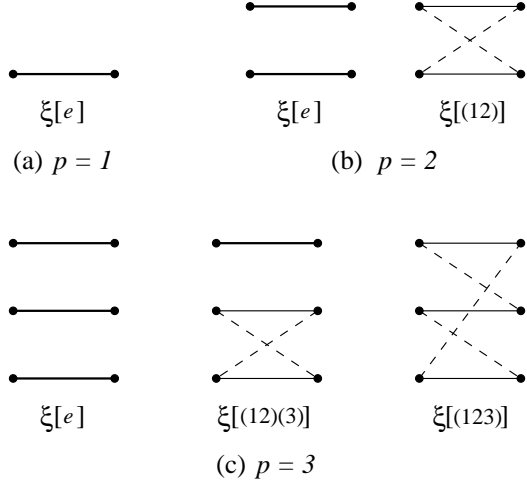}  
\caption{Primitive diagrams for (a) $p=1$, (b) $p = 2$, and (c) $p =
3$. The identity element is everywhere denoted by $e$.}
\labels{fig:prim} 
\end{center} \end{figure}

Since $\G_I=\G_J=e$, it follows from \eq{N} that $N[c]=\delta(Q\in c)$,
where $Q$ can be any element of $S_p$.
The primitive integrals \eq{intPrim} are simply $\xi[c]$, one for each
class $c$ of $S_p$. We may therefore use an element of each cycle structure
 to label the primitive integrals, as is done in \fig{prim}.
Diagrammatically, the cycle structure is translated
into the loop structure of its diagram, as can be seen in
Fig.~\ref{fig:prim}.    
Using (\ref{xi}) along with (\ref{dUn}) and the character tables found in
Appendix~\ref{sec:char} (note that $d_f = \chi_f(e)$), the primitive
integrals for $p \le 3$ can be easily computed, and the results are
displayed in Table~\ref{table:xi}. The results for $p = 4,5$ can also
be found in Table~\ref{table:xi45} of Appendix~\ref{sec:prim45}.    

\begin{table} \begin{center} 
\begin{tabular}{|c|c|c|c|} \hline
\multicolumn{4}{|c|}{$\xi[c]$} \\ [1mm] \hline
$c$ & $p=1$ & $p=2$ & $p=3$  \\ [1 mm] \hline\hline & & & \\ 
$e$ & $\frac{1}{n}$ & $\frac{1}{n^2-1}$ & $\frac{n^2-2}{n(n^2-1)(n^2-4)}$
\\ [2mm]
$(12)$ &  & $\frac{-1}{n(n^2-1)}$ & $\frac{-1}{(n^2-1)(n^2-4)}$  \\ [2mm] 
$(123)$ &  &  & $\frac{2}{n(n^2-1)(n^2-4)}$ \\ [2mm] \hline
\end{tabular}
\caption{Algebraic expressions for the primitive diagrams of $p=1,2,3$.}
\labels{table:xi}  
\end{center} \end{table}  
   
\subsection{Stack Integrals} \labels{sec:stack}

The stack diagrams (see Fig.~\ref{fig:stack}) are direct integrals made up of 
disconnected lines of arbitrary multiplicities. As such, $Q=e$, and $J$
differs from $I$ only by relabelling. Using item~3 of Sec.~2.2, we may
assume $J=I$. Hence stack integrals are integrals of the form
$\langle II|II \rangle$.  

Let $p_1, p_2,\ldots, p_t$ be the multiplicities of the disconnected lines
in a stack diagram. Then ${\cal G}_I = {\cal G}_J \equiv {\cal G} =
S_{p_1} \otimes S_{p_2} \otimes \cdots \otimes S_{p_t}$, and
$N[c]$ is nonzero only when the class $c$ is a direct product
of the classes $c_i$ of the groups $S_{p_i}$. In that case,
\be
N[c] = \prod_{i=1}^t p_i!\, n_i(c_i), \labels{Nstack}
\ee
where $n_i(c_i)$ is the number of elements of $S_{p_i}$ in the class $c_i$.
In other words, 
\be
n_i(c_i) = {p_i! \over \prod_{j=1}^{p_i} j^{\alpha_j}\alpha_j!}\ , 
\ee
where the class $c_i$ consists of $\alpha_j$ cycles of length $j$.
Denoting the stack integral $\langle II|II \rangle$ by
$\Xi(p_1,p_2,\ldots, p_t)$, we get
\begin{equation}
\Xi(p_1, p_2, \ldots, p_t) =
\sum_{c_1,c_2,\ldots} \left( \prod_{i=1}^t p_i!\, n_i(c_i) \right)
\xi(c_1\otimes c_2\otimes\cdots\otimes c_t). \labels{stack}
\end{equation}

\begin{figure} \begin{center}
\includegraphics[scale=0.9]{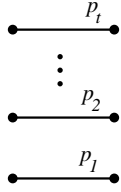}  
\caption{Arbitrary stack diagram, $\Xi(p_1, p_2, \ldots, p_t)$, of
degree $p = \sum_{i=1}^t p_i$.}  
\labels{fig:stack} 
\end{center} \end{figure}

All stack diagrams can be obtained by making the assignment $f_i
\rightarrow p_i$ from each representation. In this way, we expect a
same number of stack digrams as of primitive diagrams, or classes.
Using the $\xi$ expressions obtained in the preceding subsection, the stack
integrals for $p \le 3$ can be computed to yield the expressions
in Table~\ref{table:stack}.    

\begin{table} 
\begin{center} \begin{tabular}{|l|l|} \hline
\multicolumn{2}{|c|}{$\Xi(p_1, \ldots, p_p)$} \\ [1mm] \hline\hline  & \\
$\Xi(1) = \frac{1}{n}$ & $\Xi(3) = \frac{3!}{n(n+1)(n+2)}$ \\ [2mm]
\cline{1-1} & \\ 
$\Xi(2) = \frac{2}{n(n+1)}$ & $\Xi(2,1) = \frac{2}{(n-1)n(n+2)}$ \\ [2mm] 
$\Xi(1,1) = \frac{1}{n^2-1}$ & $\Xi(1,1,1) =
\frac{n^2-2}{n(n^2-1)(n^2-4)}$ \\ [2mm] \hline
\end{tabular}  
\caption{Algebraic expressions for the stack diagrams of $p=
1$, $2$, and $3$.} \labels{table:stack} 
\end{center} 
\end{table}

\subsection{Special Double-Fan Integrals}
The index sets for the fully opened double-fan integrals
$[A_a]^\alpha[A_b]^\alpha$  (\fig{compound2}(b) with $N=2\alpha$)
are
\be
\pmatrix{label\cr I\cr J\cr J_Q\cr}&=&
\pmatrix{1&\cdots&\alpha&\alpha+1&\dots&2\alpha\cr
b&\cdots&b&a&\cdots&a\cr
c_1&\cdots&c_\alpha&c_{\alpha+1}&\cdots&c_{2\alpha}\cr
c_{\alpha+1}&\cdots&c_{2\alpha}&c_1&\cdots&c_\alpha\cr}.
\labels{tab5}\ee
Hence both $\G_J$ and $\G_{J_Q}$ consist only of the identity $e$. As for
$\G_I$, it is given by $S_\alpha \otimes S_\alpha$, where the permutation
groups $S_\alpha$ act respectively on the $b$ and $a$ indices in
$I$. The element $Q$ maps $J_Q$ to $J$, i.e.\ $Q=(1, \alpha+1)
(2,\alpha+2) \cdots (\alpha, 2\alpha)$.  

The fully opened integral can be computed using \eq{intPrim}, with
$N[c]$ given by \eq{N} or \eq{Nq}. Thus,
\begin{eqnarray}
N[c] &=& \sum_{R\in \G_I} \sum_{T\in \G_{J_Q}}\delta(QTR \in c) \nn \\
     &=& \sum_{R\in \G_I} \delta(QR \in c)
     \;\;=\;\; \sum_{Q'} \delta(Q' \in c) \labels{sumQ},
\end{eqnarray} 
where the last sum is over {\it every} permutation $Q'$ that
sends {\it all} $b$ indices in \eq{tab5} to the positions labelled
from $\alpha + 1$ to $2\alpha$, and similarly {\it all} $a$ indices to
the positions labelled from $1$ to $\alpha$. As a consequence, the
allowed cycles of $Q'$ must be of even length, and they can be
specified by a sequence of nonnegative integers $(k) \equiv (k_1k_2
\cdots k_\alpha)$, $k_i$ being the number of cycles of length
$2i$. The number of $Q'$ with the class structure $(k)$ that is
related to $c$ is given by
\begin{equation}
N[c] = {(\alpha!)^2 \over \prod_{i=1}^\alpha i^{k_i} \cdot k_i!}.
\labels{numQ} 
\end{equation} 
In order to see how this is arrived at, consider an example where $k_1
=2$, $k_2 = 2$, and all other $k_i$ values are zero. Then $Q'$ is of
the form ($ba$)($ba$)($baba$)($baba$), where the $b$ and $a$ letters
should take the distinct index labels in $(1,\ldots,\alpha)$ and
$(\alpha+1, \ldots, 2\alpha)$ respectively. Another $Q'$ with the same
cycle structure can thus be obtained by permuting individually all the $a$
and $b$ labels. This accounts for the numerator in \eq{numQ}. However,
such permutations do not necessarily give distinct $Q'$ elements. The
cyclic nature of a cycle tells us that each cycle of length
$2i$ will appear $i$ times; this accouts for the $i^{k_i}$ factor in the
denominator. Moreover, no new $Q'$ is obtained if we permute cycles of
the same length; that accounts for the other factor $k_i!$ in the
denominator.    

We may now return to (\ref{intPrim}) to calculate the  integral
$[A_a]^\alpha[A_b]^\alpha$ in terms of the primitive integrals
$\xi[c]$. The result for the first few $\alpha$ values are listed in
Table~\ref{table:monomial}.

\begin{table} 
\begin{center} \begin{tabular}{|c|c|} \hline
$\alpha$ & $[A_a]^{\alpha}\,[A_b]^{\alpha}$ \\ [1mm] \hline\hline
  & \\
1 & $\frac{-1}{n(n^2-1)}$ \\ [1mm]
2 & $\frac{2}{(n^2-1)n^2(n+2)(n+3)}$ \\ [1mm]
3 & $\frac{-6}{(n-1)n^2(n+1)^2(n+2)(n+3)(n+4)(n+5)}$ \\ [1mm] \hline
\end{tabular}  
\caption{Values of the monomial integrals $[A_a]^{\alpha}
[A_b]^{\alpha}$ for $\alpha = 1,2,3$.}
\labels{table:monomial}   
\end{center} 
\end{table}

\begin{figure} \begin{center}
\includegraphics[scale=1]{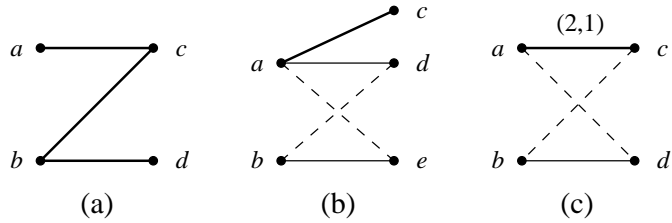}  
\caption{Non-orderly integrals of $p=3$.}
\labels{fig:diag3c} 
\end{center} \end{figure}

\subsection{Non-Orderly Integrals}
All integrals with degree $p<3$ are orderly. The non-orderly integrals
of $p=3$ are shown in \fig{diag3c}, and those related to them by
the fan relation \eq{fan}. The calculation of $N[c]$ and the integral
for each of them is discussed below. The integrals will be labelled
by their figure, {\it e.g.,} integral $I(\ref{fig:diag3c}a)$.

The index sets for \fig{diag3c}(a) are

\be
\pmatrix{label\cr I\cr J\cr J_Q\cr}=
\pmatrix{1&2&3\cr
b&b&a\cr
d&c&c\cr
d&c&c\cr}.
\labels{tab8}\ee
They give rise to the symmetry groups 
${\cal G}_I = \{ e,(12)\}$ and $\G_J={\cal
G}_{J_Q} = \{ e, (23)\}$. Moreover, the element $Q$ can be taken to be
the identity element. In order to obtain the coefficients $N[c]$
of equation (\ref{intPrim}), we need to compute $QTR$ for all $T\in {\cal
G}_{J_Q}$ and $R\in {\cal G}_I$. That triple product is  $Q{\cal G}_{J_Q}{\cal
G}_I = \{ e,(12),(23),(132) \}$. As a result, 
\be 
I(\ref{fig:diag3c}a) =Z(1,1,1)= \xi[e] + 2 \xi[{(12)(3)}] + \xi[{(123)}]
= \frac{1}{(n^2-1)(n+2)}.\labels{z111}
\ee

In the same way, the index sets of  \fig{diag3c}(b)  are
\be
\pmatrix{label\cr I\cr J\cr J_Q}=\pmatrix{1&2&3\cr
b&a&a\cr
e&d&c\cr
d&e&c\cr},\labels{tab9}\ee
and hence the symmetry groups are 
${\cal
G}_{J_Q} = \{e\}$ and ${\cal G}_I = \{ e, (23)\}$, and the exchange
element is $Q=(12)$. We thus obtain $Q{\cal G}_{J_Q}{\cal G}_I =
\{(12),(123)\}$, from which 
\be
I(\ref{fig:diag3c}b) = \xi[{(12)(3)}] + \xi[{(123)}] =
\frac{-1}{(n^2-1)n(n+2)}     
\ee
follows. 

Finally, for \fig{diag3c}(c), the index sets are
\be
\pmatrix{label\cr I\cr J\cr J_Q}=\pmatrix{1&2&3\cr
b&a&a\cr
d&c&c\cr
c&d&c\cr}.\labels{tab10}\ee
The relevant symmetry groups are
${\cal G}_{J_Q} = \{ e, (13)\}$ and ${\cal
G}_I = \{ e, (23)\}$. With $Q=(12)$, the
set $Q{\cal G}_{J_Q}{\cal G}_I$ is $\{ (12),(13),(123),(132) \}$, and
formula (\ref{intPrim}) gives:  
\be
I(\ref{fig:diag3c}c) = 2( \xi[(12)(3)] + \xi[(123)]) =
\frac{-2}{(n^2-1)n(n+2)}. 
\ee

The calculation of $Q\G_{J_Q}\G_I$ is not that cumbersome for $p=3$,
but it gets worse pretty quickly as $p$ increases.
For example, let us look at some examples of $p=4$.

Let us first calculate $Z(2,1,1)$ of \fig{excZ}(b), whose index sets are
\be
\pmatrix{label\cr I\cr J\cr J_Q}=\pmatrix{1&2&3&4\cr
b&b&b&a\cr
d&d&c&c\cr
d&d&c&c\cr}.\labels{tab11}\ee
Then $Q=e$, 
${\cal G}_{J_Q} =$
\{$e$, (12), (34), (12)(34)\} and ${\cal G}_I =$ \{$e$, (12), (13), (23),
(123), (132)\}. Thus $Q{\cal G}_{J_Q}{\cal G}_I =$ \{$e$, (12),
(13), (23), (123), (132), (12), $e$, (132), (123), (23), (13), (34),
(12)(34), (143), (243), (1243), (1432), (12)(34), (34), (1432),
(1243), (243), (143)\}, hence
\begin{eqnarray}
Z(2,1,1)\!\! &=&\!\! 2\xi[e] + 8\xi[{(12)(3)(4)}] + 8\xi[{(123)(4)}] +
2\xi[{(12)(34)}] + 4\xi[{(1234)}] \nn\\
\!\! &=&\!\! \frac{2}{(n-1)n(n+2)(n+3)}.\labels{z211}
\end{eqnarray}

\begin{figure} \begin{center}
\includegraphics[scale=0.8]{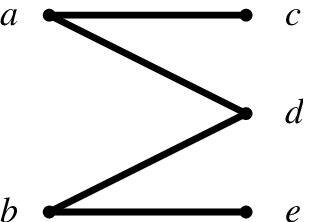}  
\caption{Sigma integral with all lines being of multiplicity one.}
\labels{fig:sigma} 
\end{center} \end{figure}

Our last example is the $\Sigma$-integral, shown in  \fig{sigma}. Its
index sets are 
\be
\pmatrix{label\cr I\cr J\cr J_Q}=\pmatrix{1&2&3&4\cr
b&b&a&a\cr
e&d&d&c\cr
e&d&d&c\cr}.\labels{tab11}\ee
Hence $Q=e$, 
 ${\cal G}_{J_Q} = \{e, (23)\}$ and ${\cal G}_I =$ \{$e$,
(12), (34), (12)(34)\}. Multiplying such elements accordingly, the
set \{$e$, (12), (34), (12)(34), (23), (132), (234), (1342)\} is
obtained for $Q{\cal G}_{J_Q}{\cal G}_I$. Hence
\begin{eqnarray}
\Sigma &=& \xi[e] + 3\xi[(12)(3)(4)] + 2\xi[{(123)(4)}] +
\xi[{(12)(34)}] + \xi[{(1234)}] \nn\\
 &=& \frac{n+1}{(n-1)n^2(n+2)(n+3)}.\labels{sigma}
\end{eqnarray}

\section{Comparison of the IM and the GTM} \label{section:comparison}
We have discussed the computation of $U(n)$ integrals \eq{intGeneral2}
in two ways: the IM in Sec.~2, and the GTM in Sec.~3. Each of these
two methods has its own merits, and drawbacks, and in a way they complement
each other. The purpose of this section is to compare their relative
stong and weak points.

The IM is based solely on the unitarity condition \eq{unitarity}, 
and the invariance of the Haar Measure \eq{inv}. The method is simple because
there is no need to know group theory. The conditions relate integrals
of the same degree, and also integrals of degree $p$ to integrals of degree
$p-1$. Through these relations, specific integrals such as the fan integrals
\eq{zf} and the $Z$-integrals \eq{Z} can be obtained, and general relation such
as the single-fan relation \eq{fan} and the double-fan relation \eq{VV}
can be worked out.

The GTM has the advantage of being general, in the sense that
all integrals can be computed using the formula \eq{intGT}
or \eq{intPrim}. The price to pay is that we have to know the characters
of the irreducible representations of the appropriate symmetric group, and
a triple sum has to be carried out, which can prove to be very tedious
for integrals of high degrees. Furthermore, unlike the IM, relations between
integrals are hard to come by, so one must calculate the integrals one by one.
 There are however certain class of integrals, the orderly
integrals, for which relations can be developed, and the quantity $N[c]$ in
\eq{intPrim} can be relatively easily computed. Then we merely have to know the
primitive integrals $\xi[c]$ in \eq{intPrim} to get the value of the orderly
integral on hand. The stack integrals \eq{stack} and the special opened double-fan
integrals \eq{numQ} are examples of this kind.
The primitive integrals
$\xi[c]$ themselves must be calculated using \eq{xi}.

To summarize, the IM gives a huge number of relations but it is not
easy to obtain the value of any specific integral. The GTM allows us to
calculate any specific integral, albeit rather tedious at times, but
it is difficult to obtain relations between integrals. In the next
section, we shall discuss a {\it hybrid method} which makes use of the
advantages of both methods. We shall use the general GTM formula to calculate
a specific set of integrals, and then use the IM relations to obtain
all the other integrals.

In the rest of this section, we shall enlarge these general remarks about the
IM and GTM, by using specific examples presented in the last two sections as
concrete illustrations.

The single-fan relation \eq{fan} can be obtained by both the IM and the GTM. However
the double-fan relation \eq{VV} in its general form can be obtained only by the IM,
because most of the integrals involved are not orderly, making it hard to derive
relations using the GTM. Nevertheless, in special cases involving only orderly
integrals, \eq{fansd}, the GTM can also be used to derive the relation.

The $Z$-formula \eq{Z} is obtained using the IM, by a series of relations
connecting it down to $\int dU=1$. Since the $Z$-integrals are not orderly, it is hard
to compute them using the GTM except at low degrees. The calculation of those
by the GTM is shown in equations \eq{z111} and \eq{z211}.

\begin{figure} \begin{center}
\epsfig{file = 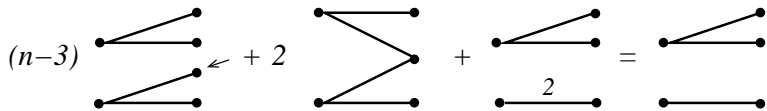}  
\caption{Unitarity sum relation involving the $\Sigma$
diagram (second from the left). The sum is performed on the index
indicated by an arrow. Using the fan relation (\ref{fan}), the
unitary sum can be written as: $(\frac{n-3}{4} + \frac{1}{2})
Z(2,0,2) + 2\Sigma = \frac{1}{2} Z(2,0,1)$.}       
\labels{fig:uniSigma} 
\end{center} \end{figure}

However, since the values 
of the integrals in the IM are obtained only through relations, it may
be relatively complicated to calculate just one specific
integral. This is where the GTM is superior, because of the general
formula \eq{intGT} valid for any one integral. For example, it is easy
to obtain the $\Sigma$ integral \eq{sigma}, assuming of course the
$\xi[c]$'s to be already known. We can also obtain it using the IM, as
we shall show below, but that involves a few steps because we must get
it from relations. To see how that is done, look at \fig{uniSigma},
which is the unitarity relation applied to the dot of the first
diagram indicated by an arrow. The first and third diagrams can be
related to $Z(2,0,2)$ by using the fan relation \eq{fan}, and
similarly the  diagram on the right can be related to
$Z(2,0,1)$. Using the $Z$-formula \eq{Z}, we then obtain 
\be
\Sigma&=&{1\over 4}\left[Z(2,0,1)-\({n-3\over 2}+1\)Z(2,0,2)\right]\nn\\
&=&{n+1\over (n-1)n^2(n+2)(n+3)},\ee
the same as the result \eq{sigma} obtained by the GTM.

\section{Hybrid Method}     
 Having understood the relative merits of the GTM and 
the IM, it is possible to combine their 
strengths into a more efficient 
hybrid calculational scheme.

The strategy is to start with one or more integrals that
can be computed by the GTM with relative ease.
Generally speaking, such integrals are ordered. Once they are
obtained, the many relations of the IM can be used to calculate other integrals
from them. 

To illustrate this strategy, we will consider how the hybrid method
can be used to calculate all double-fan integrals.

By a double-fan integral, we mean any integral with two L-dots and any number of 
R-dots. \fig{compound2}(a) shows a closed (double-fan) integral (with the 
understanding that there are no
extra dots or lines than those shown), and \fig{compound2}(b) shows a fully opened
(double-fan) integral. We may also have partially opened (double-fan) integrals,
in which every branch, namely, every R-dot with its connecting lines, can be regarded
as a closed integral.
See \fig{examples} for an example of a partially opened integral.

As in Sec.~2.2, a fully opened integral is denoted by
$[A_a]^{\alpha_a}[A_b]^{\alpha_b}[B_a]^{\beta_a}[B_b]^{\beta_b}$,
and its corresponding closed integral is denoted by 
$[\alpha_aA_a+\alpha_bA_b+\beta_aB_a+\beta_bB_b]$.
For a partially opened integral, we will denote it as
a product of the closed integrals of each branch. 
See \fig{examples} for examples.

Using \eq{VV} and \eq{v}, all double-fan integrals can be expressed
as sums of fully opened integrals. Integrals  of the form
$[A_a]^\alpha[A_b]^\alpha$  are given by \eq{numQ} and
\tab{monomial}, but we still have to know how to calculate
a fully opened integral when $\beta_i\not=0$. As shown in
Appendix~\ref{sec:openedIntReln},
the IM allows us to relate them to those with $\beta_i=0$, by 
using the following formula
\begin{eqnarray}
\lefteqn{[A_a]^{\alpha}[A_b]^{\alpha}[B_a]^{\beta_a}[B_b]^{\beta_b} =
\hspace{9mm}} \nonumber \\ 
& & \hspace{2mm} \sum_{e=0}^{\min(\beta_a,\beta_b)} \left\{
(-1)^e e! \left( \!\! \begin{array}{c} \beta_a \\ e \end{array}\!\!
\right) \left( \!\! \begin{array}{c} \beta_b \\ e \end{array}\!\! \right)
(n+2\alpha-1+2e)\; \cdot \right. \labels{monomialExpan} \\  
& & \hspace{2mm} \left.
\frac{(n+2\alpha-2+e)!\,(n+2\alpha-1+2e)!}{(n+2\alpha+\beta_a-1+e)!\,
(n+2\alpha+\beta_b-1+e)!}\: [A_a]^{\alpha + e}[A_b]^{\alpha + e}
\right\}. \nn
\end{eqnarray}

We close this section by showing how to use 
\eq{VV} and \eq{v} to calculate the integrals in 
Fig.~\ref{fig:examples}.

\begin{figure} \begin{center}
\includegraphics{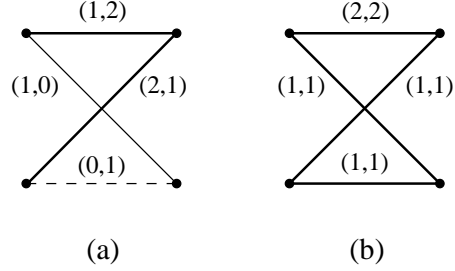}  
\caption{Partially opened double-fan integrals.
 (a) There are two equivalent forms for this graph:
$[A_a+2A_b][A_a]$and $[A_b+B_a+B_b][A_a]$. (b) 
There are four equivalent forms for this graph: $[A_a+A_b+B_a][A_a+A_b]$,\
$[2B_a+B_b][A_a+A_b]$,\ $[A_a+A_b+B_a][B_a+B_b]$,\ $[2B_a+B_b][B_a+B_b]$.}
 \labels{fig:examples}  
\end{center} \end{figure}

\subsection{\fig{examples}(a)}
There are two equivalent forms for this diagram. One is 
\be
[A_a + 2A_b][A_a]=\(2[A_a][A_b]^2\)[A_a]=2[A_a]^2[A_b]^2
\labels{10a1} \ee
and the other is
\be
[A_b+B_a+B_b][A_a]=4\([A_b][B_a][B_b]\)[A_a]=4[A_a][A_b][B_a][B_b],
\labels{10a2}\ee
where equations \eq{VV} and \eq{v} have been
used. The integral $I(\ref{fig:examples}a)$ is obtained by adding up
\eq{10a1} and \eq{10a2}.

Using \eq{monomialExpan}, we can express all fully opened integrals
in the form $[A_a]^\alpha[A_b]^\alpha$. Applying to the present case, we get
\begin{equation}
[A_a][A_b][B_a][B_b] = \frac{1}{(n+2)^2} [A_a][A_b] - \frac{1}{(n+2)}
[A_a]^2[A_b]^2. \labels{expan1}
\end{equation}      
Using \tab{monomial}, we finally obtain
\begin{eqnarray*}
I(\ref{fig:examples}a)&=& 2[A_a]^2[A_b]^2+4[A_a][A_b][B_a][B_b]\nn\\
&=&
 \frac{2n}{(n+2)} [A_a]^2[A_b]^2 +\frac{4}{(n+2)^2} [A_a][A_b] \\
 &=& \frac{-4}{(n^2-1)n(n+2)(n+3)}.
\end{eqnarray*}

\subsection{\fig{examples}(b)}
As shown in \fig{examples}(b), $I(\ref{fig:examples}b)$ has four equivalent
forms. For one branch, the factors are

\begin{eqnarray}
[A_a + A_b + B_a] & = & 4\, [A_a][A_b][B_a], \nonumber \\
\left[2 B_a + B_b\right] &=& 2\, [B_a]^2[B_b]; \label{opening2i}
\end{eqnarray}
and for the other branch, they are
 \begin{eqnarray}
[A_a + A_b] &=& [A_a][A_b], \nn \\
\left[ B_a+B_b \right]&=& [B_a][B_b]. \labels{opening2ii}
\end{eqnarray}
Hence

\begin{equation}
I(\ref{fig:examples}b) = 4\, [A_a]^2[A_b]^2[B_a] + 6\,
[A_a][A_b][B_a]^2[B_b] + 2\, [B_a]^3[B_b]^2. \labels{opening2} 
\end{equation}

We will now express each of the three monomial integrals in
(\ref{opening2}) in terms of $[A_a]^{\alpha}[A_b]^{\alpha}$. First,
with respect to (\ref{monomialExpan}), $[A_a]^2[A_b]^2[B_a]$ is
characterized by $\alpha =2$, $\beta_a = 1$, and $\beta_b = 0$. The
vanishing of $\beta_b$ causes (\ref{monomialExpan}) to consist of the
single term:
\begin{equation}
[A_a]^2[A_b]^2[B_a] = \frac{1}{(n+4)}[A_a]^2[A_b]^2. \labels{expan2i}
\end{equation}
Second, $[A_a][A_b][B_a]^2[B_b]$ has $\alpha =1$, $\beta_a = 2$,
$\beta_b = 1$, and the sum in (\ref{monomialExpan}) gives: 
\begin{equation}
[A_a][A_b][B_a]^2[B_b] =
\frac{1}{(n+2)^2(n+3)}[A_a][A_b] - \frac{2}{(n+2)(n+4)}[A_a]^2[A_b]^2.
\labels{expan2ii}
\end{equation}
Finally, $[B_a]^3[B_b]^2$, having $\alpha = 0$, $\beta_a = 3$, $\beta_b
= 2$, can be expressed as
\begin{eqnarray}
[B_a]^3[B_b]^2 &=&
\frac{1}{n^2(n+1)^2(n+2)} - \frac{6}{n(n+2)^2(n+3)}[A_a][A_b] +
\nonumber \\
 & & \frac{6}{(n+1)(n+2)(n+4)} [A_a]^2[A_b]^2 \labels{expan2iii} 
\end{eqnarray}
from equation (\ref{monomialExpan}). Using the fan relation, notice
that $[B_a]^3[B_b]^2$ can also be reduced to ${1 \over 3!}{1 \over
2!}Z(3,0,2)$.

The expressions of $[A_a][A_b]$ and $[A_a]^2[A_b]^2$ in terms of $n$
have already been determined in Example~1. The final answer is
obtained by inserting (\ref{expan2i})--(\ref{expan2iii}) into
(\ref{opening2}). The result is:
\begin{eqnarray*}
[A_a + A_b + B_a][A_a + A_b] &\!\!=\!\!&
{2 \over n^2(n+1)^2(n+2)} + {6 (n-2) \over n(n+2)^2(n+3)} [A_a][A_b] + \\
 &\!\! & {4 (n^2+2) \over (n+1)(n+2)(n+4)} [A_a]^2[A_b]^2 \\
 &\!\!=\!\!& {2 (n^2 + 2n + 4) \over (n^2-1)n^2(n+2)(n+3)(n+4)},
\end{eqnarray*}  
which can be verified using the plain group theoretical formula
(\ref{intGT}).  

\section{Conclusion}
In this article, we have pursued the goal of finding an efficient method to calculate
the monomial integral \eq{intGeneral} or \eq{intGeneral2}. We find that the IM discussed in Sec.~2 is
superior for deriving relations between integrals, but the GTM is able to give
a formula  to calculate any integral. The GTM formula involves a triple sum 
whose computation is often tedious and prone to mistakes.
The sums simplify for orderly integrals, in which
the invariant groups $\G_I$ and $\G_{J_Q}$
are either disjoint, or one is contained in the other. 
For non-orderly integrals, the hybrid method is probably
the most efficient. It uses the IM to relate them
to some orderly integrals that can be calculated by the GTM with relative ease.

This research is supported by the Natural Sciences and Engineering Research
Council of Canada and by the Fonds de recherche sur la nature et les
technologies of Qu\'ebec.

\section*{\center Appendices}
\appendix
\section{Derivation of Equations (\ref{VV}) and (\ref{v})}
\label{sec:eqns_VV_v} 

To prove \eq{VV} and \eq{v}, we use
the rotation technique discussed in item~\ref{item:rotation} of
Section~\ref{sec:IM}, and equation (\ref{rotation}), to spin off from
the R-dot of \fig{compound2}(a) 
a new R-dot attached to a pair of solid-dotted
lines. Depending on whether the basic
pattern of this new R-dot is $[A_a]$, $[A_b]$, $[B_a]$, or $[B_b]$, we
get the graphs shown in Fig.~\ref{fig:rotationL}(a),
\ref{fig:rotationL}(b), \ref{fig:rotationL}(c), and
\ref{fig:rotationL}(d) respectively.  

\begin{figure} \begin{center}
\epsfig{file = 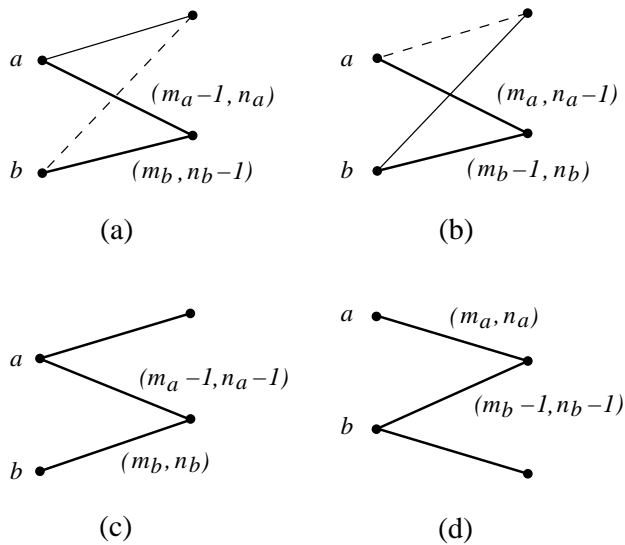}  
\caption{Diagrams that result from rotating away the pairs of lines
(a) $[A_a]$, (b) $[A_b]$, (c) $[B_a]$, and (d) $[B_b]$, from the
compound pattern of Fig.~\ref{fig:compound2}.}   
\labels{fig:rotationL} 
\end{center} \end{figure} 

Repeating this spin-off operation over and over again on the R-dot 
still containing a compound pattern, eventually we come to a graph
where every R-dot is given by a basic pattern. The resulting
integrals are given in (\ref{VV}), corresponding to the
 decompositions of the compound pattern into
a sum of all possible fully opened integrals obtained by spinning off. 

The coefficient $v(\alpha_a '\alpha_b' \beta_a' \beta_b')$ of these basic
integrals is derived from a combination of three factors: 
\begin{enumerate}

\item
Each time that we spin off a basic pattern from a compound pattern
with $d$ pairs of solid-dotted lines, there is a factor $1/d$ arising from
eq.~(\ref{rotation}), by taking $e=1$. Since we start from a compound
pattern with $N$ pair of lines, by the time we come to a fully-opened
pattern we have accumulated a factor $1/N!$.

\item
The $N$ R-dots in the final pattern that is fully opened can be
spinned off in a different sequential order. According to
(\ref{rotation}), they must be summed over. This gives rise to a factor
$N! / \alpha_a'!\alpha_b'!\beta_a'!\beta_b'!.$

\item
At any time when we spin off a basic pattern, we can choose its single
pair of solid-dotted lines in all possible ways. Eq.~(\ref{rotation})
says that we must sum over all these possibilities. The mulitipliciy
factor is given by the number of permutations of these lines that lead
back to the same basic integral. It is a factor of $m_a!\, m_b!\,
n_a!\, n_b!$.    
\end{enumerate}

Assembling these three factors, we get
\begin{eqnarray}
v(\alpha_a' \alpha_b' \beta_a' \beta_b') = {m_a!\, m_b!\, n_a!\, n_b!\,
\over \alpha_a'!\alpha_b'!\beta_a'!\beta_b'!}, 
\end{eqnarray}
which is identical to (\ref{v}).

\section{Character Tables} \labels{sec:char}
The character tables for $p=2,3$ are given here in the form used by
M.~Hamermesh in~\cite{Hamermesh}. The rows are labelled by the
partitions that define the representations, and the columns are
labelled by the cycle structures that define the classes. The number of
elements in each class, $n(c)$, is written above the classes. The
table for $p=1$ is trivial, and it consists of the sole value $1$. 

\subsection{$p=2$} \labels{sec:char2}
\begin{center} \begin{tabular}{|c|cc|} \hline
  & 1 & 1 \\
\small Part.$\backslash$Class 
  & $(1^2)$ & $(2)$ \\ [1 mm] \hline\hline 
$(2)$   & 1 & 1  \\
$(1^2)$ & 1 & -1 \\ \hline
\end{tabular} \end{center}

\subsection{$p=3$} \labels{sec:char3}
\begin{center} \begin{tabular}{|c|ccc|} \hline
  & 1 & 3 & 2 \\
\small Part.$\backslash$Class 
  & $(1^3)$ & $(1,2)$ & $(3)$ \\ [1 mm] \hline\hline
$(3)$   & 1 & 1 & 1 \\
$(2,1)$ & 2 & 0 & -1 \\ 
$(1^3)$ & 1 & -1 & 1 \\ \hline
\end{tabular} \end{center}

\vspace{1.5cm}

\section{Primitive Diagrams for $p=4$ and $p=5$} \label{sec:prim45}

Using (\ref{intGT}) and the character tables for $S_4$ and $S_5$
\cite{Hamermesh}, the algebraic expressions for the primitive
diagrams of Fig.~\ref{fig:prim4} and Fig.~\ref{fig:prim5} can be
obtained, and they are given in Table~\ref{table:xi45}. 

\vspace{1cm}

\begin{figure}[bh] \begin{center}
\epsfig{file = 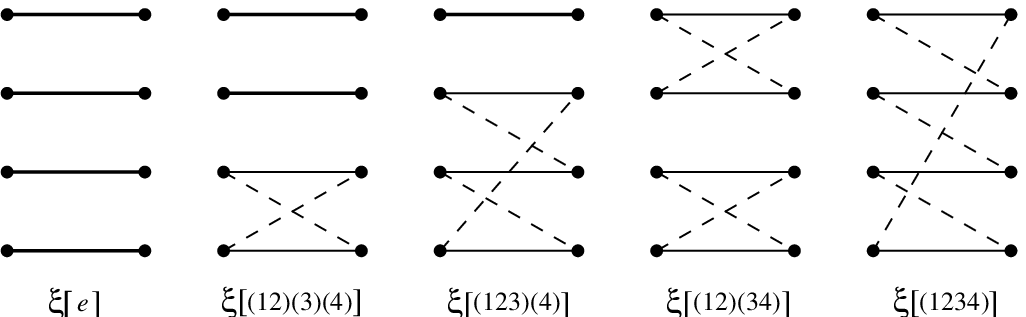}  
\caption{The $p=4$ primitive diagrams.}  
\labels{fig:prim4} 
\end{center} \end{figure} 

\begin{figure}[ht] \begin{center}
\epsfig{file = 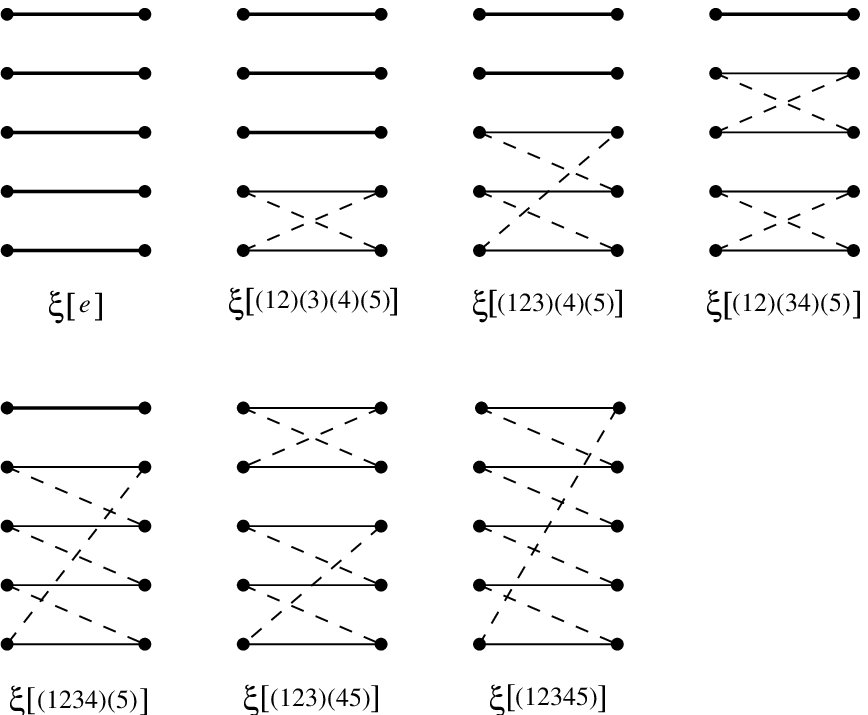}  
\caption{The $p=5$ primitive diagrams.}  
\labels{fig:prim5} 
\end{center} \end{figure}

\begin{table}[!h] \begin{center} \begin{tabular}{|c|c|c|} \hline
\multicolumn{3}{|c|}{$\xi(c_Q)$} \\ [1mm] \hline
$Q$ & $p=4$ & $p=5$ \\ [1 mm] \hline\hline & & \\ 
$e$ & $\frac{n^4 - 8n^2 + 6}{n^2(n^2-1)(n^2-4)(n^2-9)}$ & 
      $\frac{n^4 - 20n^2 + 78}{n(n^2-1)(n^2-4)(n^2-9)(n^2-16)}$ \\[2mm] 
$(12)(3)(4)$ & $\frac{-1}{n(n^2-1)(n^2-9)}$ &
            $\frac{-(n^2-2)(n^2-12)}{n^2(n^2-1)(n^2-4)(n^2-9)(n^2-16)}$
             \\ [2mm] 
$(123)(4)$ & $\frac{2n^2 -3}{n^2(n^2-1)(n^2-4)(n^2-9)}$ &
          $\frac{2}{n(n^2-1)(n^2-4)(n^2-16)}$ \\ [2mm]
$(12)(34)$ & $\frac{n^2 +6}{n^2(n^2-1)(n^2-4)(n^2-9)}$ &
             $\frac{n^2 -2}{n(n^2-1)(n^2-4)(n^2-9)(n^2-16)}$ \\ [2mm]
$(1234)$ & $\frac{-5}{n(n^2-1)(n^2-4)(n^2-9)}$ &
           $\frac{-5n^2+24}{n^2(n^2-1)(n^2-4)(n^2-9)(n^2-16)}$ \\
           [2mm]
$(123)(45)$ &  & $\frac{-2(n^2+12)}{n^2(n^2-1)(n^2-4)(n^2-9)(n^2-16)}$ \\
[2mm] 
$(12345)$ &  & $\frac{14}{n(n^2-1)(n^2-4)(n^2-9)(n^2-16)}$ \\
[2mm] \hline
\end{tabular}  
\caption{Algebraic expressions for the primitive diagrams of fourth
and fifth degrees. In the $p=5$ case, the elements from row two to
row five should be written with the additional $(5)$ one-cycle.}
\labels{table:xi45} 
\end{center} \end{table}
\vspace{4cm}

\section{Derivation of Equation~(\ref{monomialExpan})} 
\label{sec:openedIntReln}

We would like to express the general fully-opened integral
$[A_a]^{\alpha}[A_b]^{\alpha}[B_a]^{\beta_a}[B_b]^{\beta_b}$ in terms
of the special ones of the form
$[A_a]^{\alpha'}[A_b]^{\alpha'}$. The idea is to apply a
unitarity sum on the $[B_a]$ or $[B_b]$ basic patterns to get
rid of them. To get the final result we also need to apply
 the fan relation (\ref{fan}) or the double-fan relation \eq{VV} and (\ref{v}).  
Our approach is to first determine how can
$[A_a]^{\alpha}[A_b]^{\alpha}[B_a]^{\beta_a}$ be reduced to fully opened
integrals involving only the $[A_a]$ and $[A_b]$ patterns. With such an information at
hand, we will then try to  reduce the more general
$[A_a]^{\alpha}[A_b]^{\alpha}[B_a]^{\beta_a}[B_b]^{\beta_b}$ integrals
into the $[A_a]^{\alpha}[A_b]^{\alpha}[B_a]^{\beta_a}$ integrals. 

Let us apply a unitarity sum on one of the $[B_a]$ patterns in
$[A_a]^{\alpha}[A_b]^{\alpha}[B_a]^{\beta_a}$:
\begin{eqnarray}
\lefteqn{(n-(2\alpha+\beta_a-1))\:
[A_a]^{\alpha}[A_b]^{\alpha}[B_a]^{\beta_a} +
(\beta_a-1)\: [A_a]^{\alpha}[A_b]^{\alpha}[B_a]^{\beta_a-2}[2B_a]}
\nonumber \\
 & & +\: \alpha\left(\,
[A_a]^{\alpha}[A_b]^{\alpha-1}[A_b+B_a][B_a]^{\beta_a-1} +
[A_a]^{\alpha-1}[A_a+B_a][A_b]^{\alpha}[B_a]^{\beta_a-1}\, \right) 
\nonumber \\
 & & = [A_a]^{\alpha}[A_b]^{\alpha}[B_a]^{\beta_a-1} \labels{uniSum1}
\end{eqnarray}
Using (\ref{fan}), $[2B_a]$, in the second term above, can be
rewritten as $2[B_a]^2$. Furthermore, equation (\ref{v}) tells us that
$[A_b + B_a]$ and $[A_a + B_a]$ can respectively be rewritten as
$2[A_b][B_a]$ and $2[A_a][B_a]$, and the term in parentheses
above simplifies to
$4[A_a]^{\alpha}[A_b]^{\alpha}[B_a]^{\beta_a}$. As a result, relation
(\ref{uniSum1}) reduces to:
\begin{equation}
[A_a]^\alpha[A_b]^\alpha[B_a]^{\beta_a} = {1 \over n + 2\alpha +
\beta_a -1}\, [A_a]^\alpha[A_b]^\alpha[B_a]^{\beta_a - 1}. \labels{uniSum1ii}
\end{equation}
Using the relation (\ref{uniSum1ii}) recursively on its right-side,
until no $[B_a]$ remains, we obtain: 
\begin{eqnarray}
\lefteqn{[A_a]^\alpha[A_b]^\alpha[B_a]^{\beta_a}} \nonumber \\
 &= & {1 \over (n + 2\alpha + \beta_a -1)} {1 \over (n + 2\alpha +
 \beta_a -2)} \cdots {1 \over (n + 2\alpha)}\,
 [A_a]^\alpha[A_b]^\alpha \nonumber \nonumber \\
 &=& {(n + 2\alpha -1)! \over (n + 2\alpha + \beta_a -1)!}
 \,[A_a]^\alpha[A_b]^\alpha, \labels{uniSum1iii}
\end{eqnarray}
and the first step of the work is completed.

Assuming that $\beta_b \leq \beta_a$, let us perform a unitarity sum on a
$[B_b]$ pattern in
$[A_a]^\alpha[A_b]^\alpha[B_a]^{\beta_a}[B_b]^{\beta_b}$:  
\begin{eqnarray}
\lefteqn{(n-(2\alpha+\beta_a+\beta_b-1))\:
 [A_a]^\alpha[A_b]^\alpha[B_a]^{\beta_a}[B_b]^{\beta_b}} \nonumber \\ 
&+&
 (\beta_b-1)\:[A_a]^\alpha[A_b]^\alpha[B_a]^{\beta_a}[B_b]^{\beta_b-2}[2B_b] 
 \nonumber \\ 
&+& \beta_a\: [A_a]^\alpha[A_b]^\alpha[B_a]^{\beta_a-1}[B_a +
 B_b][B_b]^{\beta_b-1} \nonumber \\  
&+& \alpha \left(\,
 [A_a]^{\alpha}[A_b]^{\alpha-1}[A_b+B_b][B_a]^{\beta_a}[B_b]^{\beta_b-1}
 \right. \nonumber \\
& & \left. \hspace{5mm} +\;
 [A_a]^{\alpha-1}[A_a+B_b][A_b]^{\alpha}[B_a]^{\beta_a}[B_b]^{\beta_b-1}\, 
 \right) \nonumber \\
 &=& [A_a]^{\alpha}[A_b]^{\alpha}[B_a]^{\beta_a}[B_b]^{\beta_b-1}.
 \labels{uniSum2} 
\end{eqnarray}
Relation (\ref{fan}), or formula~(\ref{v}), again permit to make
some simplifications, i.e.\ $[2B_b] = 2[B_b]^2$, $[B_a + B_b] =
[A_a][A_b] + [B_a][B_b]$, $[A_b + B_b] = 2[A_b][B_b]$, $[A_a + B_b] =
2[A_a][B_b]$. By making the proper substitutions in (\ref{uniSum2}), the
recursion equation, 
\begin{eqnarray*}
\lefteqn{[A_a]^\alpha[A_b]^\alpha[B_a]^{\beta_a}[B_b]^{\beta_b} = } \\
 & & {1 \over (n+2\alpha+\beta_b-1)} \left\{
[A_a]^\alpha[A_b]^\alpha[B_a]^{\beta_a}[B_b]^{\beta_b-1}
\right. \\
 & & -\: \left. \beta_a\,
[A_a]^{\alpha+1}[A_b]^{\alpha+1}[B_a]^{\beta_a-1}[B_b]^{\beta_b-1}  
\right\}, 
\end{eqnarray*}
results. The above can be solved to give
\begin{eqnarray*}
\lefteqn{[A_a]^{\alpha}[A_b]^{\alpha}[B_a]^{\beta_a}[B_b]^{\beta_b} = } 
\\ 
& & \sum_{e=0}^{\beta_b} \left\{
(-1)^e e! \left( \!\! \begin{array}{c} \beta_a \\ e \end{array}\!\!
\right) \left( \!\! \begin{array}{c} \beta_b \\ e \end{array}\!\! \right)
(n+2\alpha-1+2e)\; \cdot \right. \\  
& & \left.
\frac{(n+2\alpha-2+e)!}{(n+2\alpha+\beta_b-1+e)!}\: [A_a]^{\alpha +
e}[A_b]^{\alpha + e}[B_a]^{\beta_a-e} \right\},
\end{eqnarray*}
which upon substitution of (\ref{uniSum1iii}) yields the desired equation
(\ref{monomialExpan}).  


\end{document}